\documentstyle[epsf]{l-aa}

\def\Ra{\left(\frac Ra\right)}
\def\wwcr{\left(\frac{\omega}{\omega_{cr}}\right)}
\def\vc{\left(\frac vc\right)}
\def\lam{\left(\frac{L_{em}}{L_{GR}} \right)}
\def\dddot#1{\stackrel{\dots}{#1}}

\begin{document}
\title{Pulsars revived by gravitational waves }

\author{Vladimir M. Lipunov$^{1,2}$ \and
       Ivan E. Panchenko  $^2$
       }
\date{}
\institute{Department of Physics, Moscow State University \and
Sternberg Astronomical Institute,   13, Universitetskij pr., Moscow, Russia}
\maketitle

\begin{abstract}
 Binary neutron stars mergers that are expected to be the most powerful source of
 energy in the Universe definitely exist in nature, as is proven
 by the observed behavior of the Hulse-Taylor binary radio pulsar
(\cite{HulseTaylor}).
 Though most of energy in such events is radiated in gravitational
 waves, there probably exist several mechanisms giving also electromagnetic
 radiation. We propose a new one, involving a revival of the radio pulsar
 several orbital cycles before the merger.
\end{abstract}

\section{Introduction}
Observations of binary radio pulsars (fast rotating neutron stars)
made it possible to determine that the orbits of binaries contract in exact
correspondence with the Einstein's formula for emission of gravitational
waves
$$  L_{gw} = \frac{4}{5} \frac{G^4}{c^5} \frac{M^5}{a^5} $$
where $M$ is the mass of the components, $a$ is the semimajor axis of the
binary, $G$ and $c$  are the gravitational constant and velocity of light,
respectively.
Moreover, in many cases the characteristic time for this orbit contraction
is less than the age of Galaxy, which implies that the neutron star mergers
in the Universe should necessarily occur (\cite{HulseTaylor}).

The merger, if the stars are compact, is inevitably
accompanied by energy output of the order of $Mc^2$ in a very small
time of the order of $R_g/c$, so the power of this process is close to the
maximal possible value $L_{GR} = E_{Plank}/t_{Plank} = c^5/G \sim 10^{59}$~erg/s
(\cite{Lip93}).
Of course, much of this energy is carried away by gravitational
waves that may be soon detected by the new laser
interferometers (\cite{ligo}) that are now under construction or under
consideration (\cite{lisa}), but probably some part will be radiated as
electromagnetic waves.

The analysis of the modern scenario of evolution of binaries provides the
rate of the neutron star merger events in the Galaxy of
once per $10^4-10^6$ years (\cite{LipE}, \cite{tutuk}, \cite{vdH},
Lipunov et al, 1995,
Portegies Zwart, Spreeuw, 1995 ) for
various distributions of initial parameters.
The least estimate ($1$ per $10^6$ years) of this rate is given by
Phinney (1991) on the basis of the observed population of
binary pulsars .

Though the value of the merger rate is not significant for existence of the
proposed mechanism of radio revival, we will make the estimations of
its observational properties on the basis of the most optimistic estimate
of the rate: $10^{-4}$~yr$^{-1}$.

As the total mass of all galaxies in the Universe is approximately
$10^{9}$ times greater than the mass of our Galaxy,
such event happens approximately
once per minute. The enormous energy released in this process makes it
attractive for explanation of gamma-ray bursts (\cite{blin,PacA,PacB}),
but though it is still not
clear what part of energy is transformed into electromagnetic radiation and what
is the mechanism of such transformation.

In this paper we show that during the inspiral of two neutron stars
that precedes their coalescence if
at least one of them has a strong magnetic field,
the conditions for revival of the classical pulsar mechanism (\cite{gold,GJ})
would necessarily arise
due to fast orbital motion.

\section{The Mechanism for the Pulsar Revival}
Let us remind that the energy output of the pulsars is caused by the spinning
of a dipole magnetic field and is well described by the classical formula for
the magnetic dipole emission (\cite{Landau}):

$$   L_m = \frac{2}{3c^3} \mu^2 \omega^4   $$
where $\mu$ is the magnetic dipole momentum and $\omega$ is the spin
angular velocity.

Now we observe about $600$ radio pulsars loosing their spin energy
in accordance with this formula. Usually their luminosity does not
exceed several units of the luminosity of the Crab pulsar
($10^{38}$~erg s$^{-1}$). The reason is that pulsars are evidently born
with relatively long period of the order $10^{-2}$~sec
while the magnetic dipole momentum is $\mu\approx 10^{30}$~G cm$^3$. So
the standard pulsars have rather low luminosity and in several million
years spin down and die.

The fate of a neutron star in a close binary system is quite different.
During the accretion phase the neutron star is spun up to periods close
to the critical ($T_{cr}=2\pi/\omega_{cr}\sim 1$~ms), what leads to the revival of the pulsar mechanism
after the accretion stops even if the magnetic field would be reduced by
that time.

During the inspiral the orbital period of the binary decreases reaching the
minimum value of $1$~ms  when the components come into mechanical contact.

In all further calculations we assume the masses of both neutron stars
to be equal to $M=1.4M_\odot$; the orbit to
be circular with a separation between the components $a$. The neutron
star radius $R$ is taken $10$~km.

\subsection{The Synchronized Dipole}
Only for the purpose of comparison,
assume the really impossible case if the neutron star spin is synchronized
with the orbital motion. Then $\omega_{spin} = \omega_{orb}$ -- so the
neutron star will behave like a classical pulsar with a very high frequency,
emitting energy according to the magnetic dipole formula.
In practice,
due to enormous increase of $\omega_{orb}$ it
will essentially exceed the spin angular
velocity $\omega_{spin}\sim 0.1\ldots 10 $~s$^{-1}$
after some moment of time,
and the tidal forces will not be strong enough to spin the neutron star up
to $\omega_{orb}$, so synchronization will be impossible.

It is so because the tidal relaxation time
that can be derived from observations of pulsar glitches is of the order
of several days which is much greater than the lifetime of the binary system
with such orbital period.

Thus, below we can assume
$\omega_{spin} \ll  \omega_{orb} $, or for crude estimate, $\omega_{spin}=0$.

\subsection{The Orbital Quadrupole}
If a magnetic dipole $\mu$ performs orbital motion say, on a circular orbit
of a radius $a$,
the total quadrupole moment ${\bf D}$ in the inertial frame corresponding
to the center of the orbit will change as
$$
 D^2(t)= 18 \mu^2 a^2 \left(1 + \frac{\cos^2\theta(t)}{3}\right)
$$
where $\theta$ is the angle between the orbital radius and
the magnetic dipole axis, $\cos\theta(t) = \cos i \cos\alpha(t) $,
where $i$ is the dipole inclination to the orbit plane
and $\alpha$ is the phase of orbital motion.

It should cause magnetic quadrupole energy losses with rate

$$L = \frac{{\dddot D}^2}{180c^5}, $$
where $$ \dddot D = \omega_{orb}^3 a\mu $$

\subsection{The Induced dipole}
Consider the binary system containing two neutron stars with low
spin frequencies $\omega\sim 0$, one of them having significant magnetic
field.
Because of this enormous increase of $\omega_{orb}$ it
will essentially exceed the spin angular
velocity $\omega_{spin}\sim 0.1\ldots 10 $~s$^{-1}$
after some moment of time,
so we can assume
$\omega_{spin} \ll  \omega_{orb} $, or for crude estimate, $\omega_{spin}=0$.
The $\omega_{spin}$ does not significantly change during the inspiral
because the tidal forces that tend to synchronize the spin and orbital
motion are negligibly small in this case.

\begin{figure}[ht]
\epsfxsize=\hsize
\epsfbox{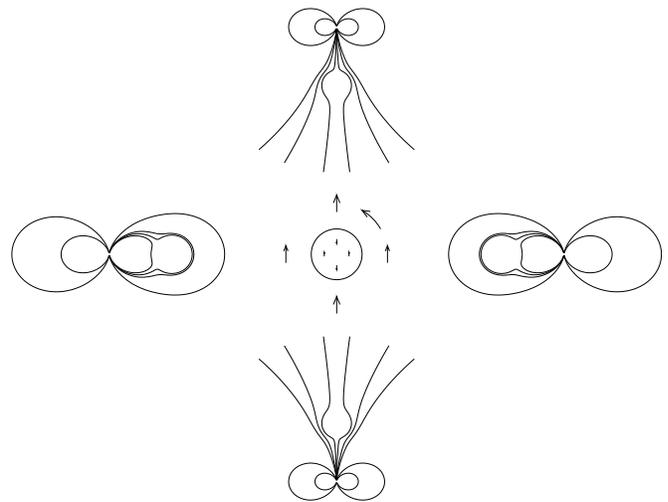}
\caption{The field configurations in $4$ positions of a magnetic dipole
(the big arrows) near a superconducting sphere. The small arrows
represent the induced dipole positions.}
\end{figure}

In order to find the field configuration, let us replace the magnetized
star by a dipole $\mu_0$ tilted to the angle $\alpha$ to the orbital axis
and solve the magnetostatic problem of a dipole
in the vicinity of a superconducting sphere. The result is (see fig.1) that
the induced magnetic field of the sphere will be equal to the magnetic
field of the dipole $\mu_1 = \mu_0 \Ra^3$ tilted oppositely to the tilt
of the outer dipole and shifted from the center of the sphere to the
distance $R^2/a$ .
If we rotate the dipole $\mu_0$ around the sphere, the induced dipole $\mu_1$
will rotate twice. This means that if the dipole is orbiting around the sphere,
the total dipole momentum
 $\mu = \mu_0 + \mu_1$ would consist of a constant part and a part oscillating
with a double orbital frequency, which implies the existing of magnetic dipole
energy loss

$$
 L_(t) = \frac{8\mu^2 \sin^2\alpha \omega_{orb}^8}{3c^3\omega_{cr}^4}
   \sim 5\cdot 10^{32}\sin^2\alpha\mu_{30}^2 t^{-3}\mbox{ erg s}^{-1}
$$
that should be carried away from the system by electromagnetic radiation
or accelerated charged particles.

\begin{figure}[ht]
\epsfxsize=6cm
\centerline{\epsfbox{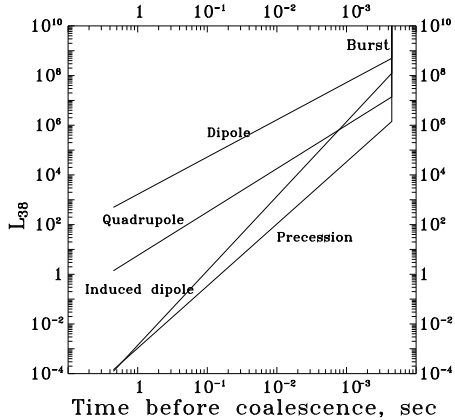}}
\caption{ The light curve of the revived pulsar with
$\mu=10^{30}$~G~cm$^3$. $L_{38}=L/10^{38}$~erg/s. }
\end{figure}

The infinite growth of luminosity (fig 2.) is in fact terminated by the
finite size of the stars. We can apply the above formula only when the
stars can be considered as two separate bodies, i.e. the distance
between them is greater than $2R$.
So the maximum luminosity reached when the stars come into surface contact
will be
$$
 L_{max} \sim 8\cdot 10^{41}\sin^2\alpha\mu_{30}^2 \mbox{ erg s}^{-1}
$$

Of course, in the case of the induced dipole the configuration of the
magnetosphere is quite different from that in the case of a solitary pulsar.
The most important condition for particle acceleration is the existence of
a vacuum region
(at least the density of the charges should be less than in Goldreich --
Julian case)
where the electric field would be not orthogonal to the magnetic field.
We point out that in our case at least near the neutral points where the magnetic
field is equal to zero we have a nonzero electric field of both relativistic
($\frac vcB$) and inductive ($\partial B/\partial t$) origin.

We emphasize that this mechanism is a modification of classical unipolar
inductor mechanism of particle acceleration. It is essentially
different from that proposed by Nulsen and Fabian (\cite{Nulsen})
for Geminga.

\subsection{The Lense-Thirring Precession}
In strong gravitational field of an orbiting binary star the
Lense-Thirring precession of the stars should occur. Its frequency
is (\cite{lensethir})
$$\Omega_g\approx  \frac 78 \frac{(GM)^{2/3}\omega_{orb}^{5/3}}{2^{1/3}c^5} $$
It leads to rotation of the magnetic dipole moment at the same frequency and
thus, magnetic dipole energy losses.

\subsection{Summary}
Thus the pre-merger, inspiral of the neutron stars should be
accompanied by all types of pulsar activity which should be displayed,
according to the observed properties of Crab-like pulsars, in a wide
range of electromagnetic spectrum from radio to gamma rays.
The characteristic feature of this pulsar activity is that the
frequency of its pulsations ($\omega=2\omega_{orb}$ for induced dipole and
$\omega=\omega_{orb}$ for other cases) should increase with
time according to the gravitational radiation orbit contraction law.

As the effect might be observed both as electromagnetic radiation and
as gravitational radiation waveform, we present a summary of both
electromagnetic luminosity $L$ and its ratio to gravitational wave luminosity
$L/L_{gw}$ in the table \ref{formulae}. In this table we
took $x=R/R_g = 3$,
$\omega_{cr} = \sqrt{GM/R^3}$,
$R_g =\sqrt{2GM/c^2}$,
$L_{em} =\frac 32 \frac{\mu^2c}{R_g^4} $ - the ``maximal'' electromagnetic
luminosity. $v/c$ is the ratio of the characteristic orbital velocity
to the velocity of light.

\def\Ld{5.0\cdot 10^{46}}
\def\ld{4.8\cdot 10^{-13}}
\def\Lq{1.4\cdot 10^{45}}
\def\lq{2.6\cdot 10^{-14}}
\def\Lp{1.4\cdot 10^{44}}
\def\lp{6.1\cdot 10^{-4}}
\def\Li{1.2\cdot 10^{46}}
\def\li{5.5\cdot 10^{-12}}

\begin{table*} \label{formulae}
\caption{The characteristics of all 4 mechanisms of pulsar revival.}
\begin{tabular}{lccc}
\hline\hline
   Mechanism & $L$ &$L$,erg/s & $L/L_{gw}$   \\                     \hline
 Dipole    & $L_{em}\left(\frac{1}{2x^3}\right)^2\wwcr^4 $
           & $\sim \Ld\wwcr^4 $
           &$\frac{1}{2^{11}}\lam \vc^2
           \sim \ld \vc^2$\\ \\
 Quadrupole&$L_{em}\frac{2^{1/3}}{60x^7}\wwcr^{14/3}$
           &$ \sim \Lq \wwcr^{14/3}$
           &$\lam\frac{1}{2^{11\frac{1}{3}}\cdot 15}\vc^4
            \sim \lq\vc^4$ \\ \\
 Induced dipole &
            $L_{em}\frac{1}{16x^6}\wwcr^8 $
           &$\sim \Li\wwcr^8 $
           &$\lam\frac{x^6}{2^{17}}\vc^{14}
           \sim \li\vc^{14}$ \\  \\
 Precession&$L_{em}\left(\frac{7}{8}\right)^4 \frac{1}{2^{10/3}x^{10}} \wwcr^{20/3} $
           &$\sim \Lp\wwcr^{20/3} $
           &$2^{20}\lam \left(\frac{7}{8}\right)^4\vc^{20}
           \sim \lq\vc^{20}$ \\ \\
\hline\hline

\end{tabular}
\end{table*}

\section{Discussion}
As the classical pulsars where first discovered in the radio wave band, we
can suppose that the most prospective for the observations of the
inspiralling neutron stars could be also radio emission.

Suppose that one of the stars has the characteristics of the Crab pulsar.
Then at the spin frequency close to the orbital one of the inspiralling
binaries ($\sim 1$~ms) its luminosity should be $10^6$ times higher.
This means that this coalescing binary will display a radio burst with a
flux equal to the flux from the Crab from distances of $2$~Mpc, and if
we take into account that the sensitivity of modern radio telescopes allows
to see pulsars $1000$ times less luminous than the Crab we can see this
bursts from the distances of $60$~Mpc. The rate of these events in
this part of the Universe is optimistically estimated as several events
per year and even with accounting for
the beam of the telescopes we can expect their observation
to be possible.

In the optical band, the revived pulsar should have the maximum absolute
magnitude $M=-9$. We can see that the problem of its detection is rather
difficult in comparison with the detection of distant supernovae.

Of course, it would be attractive to connect the proposed mechanism with the
gamma-ray bursts themselves. In fact, it is possible to suppose following
Usov (\cite{Usov}) that some fraction $\sim 1\%$ of the pulsars have
enormously strong magnetic fields ($10^{15}-10^{16}$~G), then
the pulsar mechanism can provide the luminosity required for the cosmological
gamma-ray bursts. This seems reasonable, as the total rate of the neutron star
mergers is $100$--$1000$ times higher than the rate of gamma-ray bursts
in the cosmological model. However, one should also take into account
the anisotropy of the radiation necessarily arising in the pulsar mechanism,
that is used to balance the rates of the gamma-ray bursts and neutron star
mergers (\cite{LipN}). We however note that irrespectively of the
real nature of gamma-ray bursts, the gravitational wave bursts that mainly
originate from the coalescence of the neutron stars should have the
pulsar-like precursors.

Obviously, on the last phase of the inspiral when the neutron star crust is
strongly deformed and the neutron stars themselves are destroyed --
the pulsar mechanism should be damped to some degree.

\vskip\baselineskip
\noindent
ACKNOWLEDGMENTS
We would like to acknowledge Drs G.V. Lipunova, L.P.Grischuk,
M.E. Prokhorov and
S.N. Nazin for their helpful discussions of the subject.
We are specially grateful to Dr V. Beskin who pointed to the possibility of
the orbital quadrupole effect and Dr K.A. Postnov who attracted our attention to the
Lense-Thirring precession.
The research described in this publication was made possible in part by Grant
N JAP-$100$ from the International Science Foundation and Russian Government.
The work of Ivan Panchenko has been made possible by a
fellowship of Tomalla foundation and is carried out under the research
program of International Center for Fundamental Physics in Moscow.

\end{document}